\documentclass[aps,twocolumn,superscriptaddress]{revtex4-1}
\usepackage{graphicx}
\usepackage{amsmath}
\usepackage{subfigure}
\usepackage{braket}
\usepackage{color}
\begin{document}


\title{Velocity preserving transfer between highly excited atomic states: Black Body Radiation and Collisions}


\author{J.C de Aquino Carvalho}
\affiliation{Laboratoire de Physique des Lasers, Universit{\'e} Paris 13, Sorbonne Paris-Cit{\'e}, F-93430, Villetaneuse, France}
\affiliation{CNRS, UMR 7538, LPL, 99 Avenue J.-B. Cl{\'e}ment, F-93430 Villetaneuse, France}

\author{I. Maurin }
\affiliation{Laboratoire de Physique des Lasers, Universit{\'e} Paris 13, Sorbonne Paris-Cit{\'e}, F-93430, Villetaneuse, France}
\affiliation{CNRS, UMR 7538, LPL, 99 Avenue J.-B. Cl{\'e}ment, F-93430 Villetaneuse, France}

\author{H. Failache }
\affiliation{Instituto de Fısica, Facultad de Ingenierıa, Universidad de la Republica, J. Herrera y Reissig 565, 11300 Montevideo, Uruguay}

\author{D. Bloch }
\affiliation{CNRS, UMR 7538, LPL, 99 Avenue J.-B. Cl{\'e}ment, F-93430 Villetaneuse, France}
\affiliation{Laboratoire de Physique des Lasers, Universit{\'e} Paris 13, Sorbonne Paris-Cit{\'e}, F-93430, Villetaneuse, France}

\author{A. Laliotis }
\email{laliotis@univ-paris13.fr}
\affiliation{Laboratoire de Physique des Lasers, Universit{\'e} Paris 13, Sorbonne Paris-Cit{\'e}, F-93430, Villetaneuse, France}
\affiliation{CNRS, UMR 7538, LPL, 99 Avenue J.-B. Cl{\'e}ment, F-93430 Villetaneuse, France}



\date{\today}

\begin{abstract}
We study the excitation redistribution from cesium  $7\mathrm{P}_{1/2}$ or $7\mathrm{P}_{3/2}$ to neighboring energy levels by Black Body Radiation (BBR) and inter atomic collisions using pump-probe spectroscopy inside a vapor cell. At low vapor densities we measure redistribution of the initial, velocity-selected, atomic excitation by BBR. This preserves the selected atomic velocities allowing us to perform high resolution spectroscopy of the $\mathrm{6D\rightarrow 7F}$ transitions. This transfer mechanism could also be used to perform sub-Doppler spectroscopy of the cesium highly-excited nG levels. At high densities we observe interatomic collisions redistributing the excitation within the cesium $\mathrm{7P}$ fine and hyperfine structure. We show that $\mathrm{7P}$ redistribution involves state-changing collisions that preserve the initial selection of atomic velocities. These redistribution mechanisms can be of importance for experiments probing high lying excited states in dense alkali vapor.   
\end{abstract}

%
%
%
%
%
\maketitle

\section{Introduction}
Vapor cells are attractive compact platforms for fundamental physics and quantum technology experiments. For most applications, alkali metal vapor cells are required to operate in elevated temperatures to increase the available atomic density. Under these conditions the effects of BBR and of inter-atomic collisions on the atomic population distribution and lifetime become of importance, particularly for experiments involving highly excited atomic states that are now more easily accessible due to advances in laser diode technology. 

Probing atomic vapors at elevated atomic density requires understanding of collisional mechanisms. Collision assisted velocity redistribution (thermalization) within the hyperfine manifold of the cesium $6P$ levels \cite{huennekens_pra_1995} has been measured in vapor cells and has been of importance for experiments near surfaces \cite{fichet_epl_2007,failacheprl1999,laliotisnatcommun2014}. Collisional effects have also been studied for higher lying states like the $\mathrm{7P}$  \cite{pace_canjp_1974, leslie_jap_1977, chevrollieroptlett1991} and the $\mathrm{8P}$  \cite{pimbert_1972, segundo_laserphysics_2007} levels of cesium, where indirect evidence of BBR redistribution was also reported \cite{pimbert_1972}. Nevertheless, the effects of collisions in redistributing or thermalizing the atomic velocities has remained so far unexplored for high-lying excited states for which radiation trapping and resonant exchange collisions with ground state atoms are expected to reduce. This could unmask collisional mechanisms that preserve a memory of the laser selected velocities, previously studied mainly with molecules \cite{Weber81} or buffer gas perturbers \cite{haverkort_pra_1987, Berman_PRA_1980, Morgus_pra_2008}.    

The interaction of atoms with BBR has been mainly studied for Rydberg atoms \cite{gallagher_prl_1979, spencer_pra_1982, cooke_pra_1980, oliveira_pra_2002, farley_pra_1981, beterov_pra_2009}, that have many dipole couplings at mid and far infrared (thermal) wavelengths. The depopulation of Rydberg states due to BBR has been studied experimentally and theoretically in the volume \cite{farley_pra_1981, beterov_pra_2009} or inside a cavity \cite{lai_prl_1998}. The effects of BBR have also been studied on trapped molecules for which BBR is the main thermalisation mechanism \cite{hoekstra_prl_2007}.

Our group is interested in the interaction of atoms with thermal fields in the near field of a hot surface \cite{laliotisnatcommun2014, failacheprl1999, failacheEPJD2003, gorza_EPJD_2006, laliotis_pra_2015}. Contrary to far-field BBR, near field thermal emission is monochromatic due to the thermal excitation of evanescent surface polariton modes \cite{shchegrov_prl_2000,GreffetNature2002}. The near-field redistribution of the atomic excitation is expected to display distinct characteristics that need to be discriminated and distinguished from volume redistribution due to collisions or BBR. The cesium $\mathrm{7P}$ atoms are of particular interest in such experiments due to their coupling with the $\mathrm{6D}$ states that coincides with the sapphire polariton modes \cite{failacheprl1999, failacheEPJD2003, Joao}.  


Here, we investigate far-field BBR and collision redistribution mechanisms, from cesium $\mathrm{7P}$ levels, in the volume of a sapphire cell. The unusual feature of our cell is that its main body can be heated to high temperatures up to 1000 K, while the cesium density is almost independently regulated via the temperature of the cesium reservoir. The pump laser, exciting atoms to the $\mathrm{7P}$ level, also selects the atomic velocity along the beam propagation axis. At moderate cesium densities we study velocity preserving BBR transfer to the $\mathrm{6D}$ states  by probing the $\mathrm{6D \rightarrow 7F}$ transition (see Fig.1a). Our set-up allows us to explore BBR
effects from 400 K up to 1000 K. We also study collisional redistribution within the fine and hyperfine states of the $\mathrm{7P}$ levels by probing the $\mathrm{7P \rightarrow 10S}$ transitions (see Fig.1a). Our experiments allow measurements of the velocity distribution of the excited state population. We show that radiation trapping and exchange collisions with the ground state do not efficiently thermalize velocities in the $\mathrm{7P}$ levels. Instead, we observe the existence of fine and hyperfine structure changing collisions that preserve the velocity selection. These collisions possibly involve two excited atoms whose velocities are selected by the pump laser. 

\section{Experiment and results}

A schematic of our pump-probe experimental set-up is shown in Fig.1b. Our cell is an 8cm long sapphire tube onto which two sapphire windows are attached using a high temperature mineral glue  \cite{laliotisnatcommun2014, Sarkisyan2000}. The cell was previously used to perfom atom-surface interaction measurements at high temperatures up to 1000 K \cite{laliotisnatcommun2014}. The maximum operating temperature of the first window is about 1300 K. The cell contains a sidearm, which acts as a cesium reservoir, glued around a hole drilled on the second window. The maximum operating temperature of the second window is about 700 K. 

Two independent ovens control the cell body temperature and a third oven controls the reservoir. Inside each oven, the temperature is measured by thermocouples that are in contact with the first window, $\mathrm{T_{cell}}$, second window, $\mathrm{T_{inter}}$, and and cesium reservoir, $\mathrm{T_{r}}$. The experimental error bars in these measurements are $\mathrm{\Delta T \approx}$ 30 K, mostly due to systematic uncertainties. Below 700K the temperature of the cell body ($\mathrm{T_{inter}}$ and $\mathrm{T_{cell}}$) is kept as homogeneous as possible within the experimental precision of the temperature measurements. However, for values of $\mathrm{T_{cell}}$ larger than 700 K, $\mathrm{T_{inter}}$ stays fixed at 700 K in order to protect the cell. This means that the cell body temperature is inhomogeneous. The effective temperature relevant for our measurements is often $\mathrm{T_{cell}}$, depending on the laser beam propagation inside the cell. The reservoir temperature ($\mathrm{T_r}$) is varied from 330 K to 470  K. Our studies \cite{Joao} have shown that the cesium vapor pressure inside the cell can be considered constant, defined by the reservoir temperature,$\mathrm{P(T_r)}$, while the cesium density depends on the local temperature of the cell body. When the temperature of the body is homogeneous the cesium density is given by $\mathrm{n_{Cs}=P(T_r)/k_{B}T_{cell}}$, where $\mathrm{k_{B}}$ is the Boltzmann constant.    

\begin{figure}[ht]
\begin{center}
\includegraphics[width=80mm]{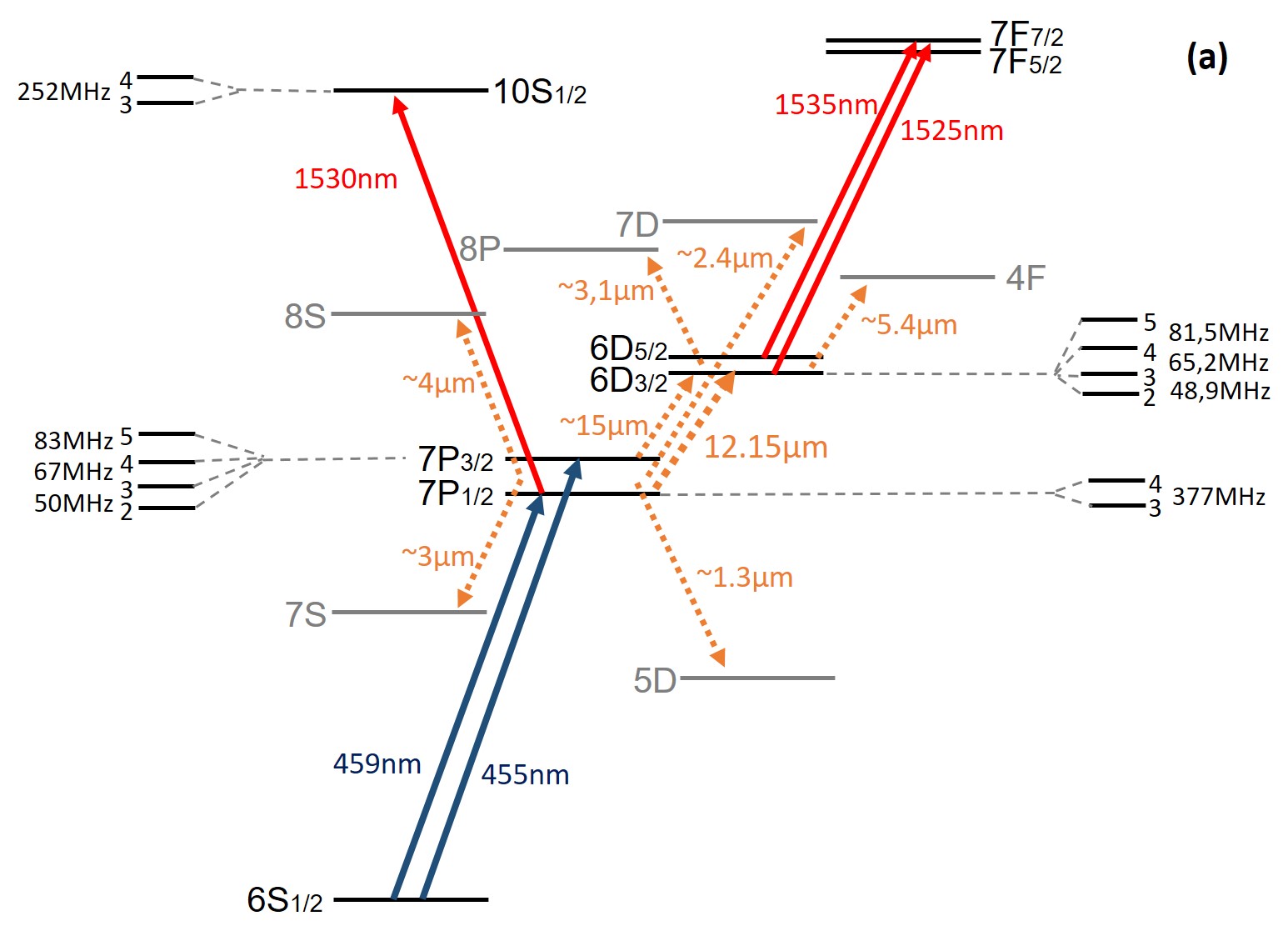}\\%
\includegraphics[width=80mm]{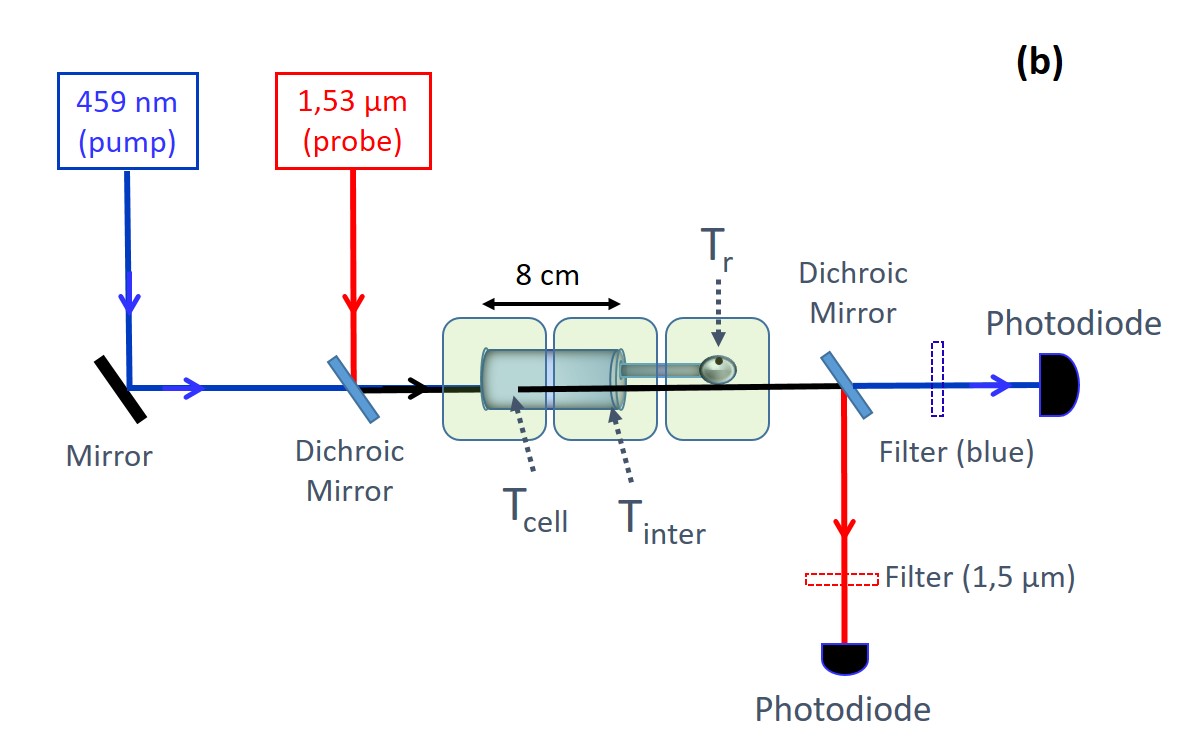}%
\end{center}
\caption{ (a) Schematic of the cesium levels relevant for this experiment. The BBR couplings relevant for this experiment can occur in the $\mathrm{7P_{1/2} \rightarrow 6D_{3/2}}$ transition at $\mathrm{12.15\;\mu m}$ and in the $\mathrm{7P_{3/2} \rightarrow 6D_{3/2}}$ and $\mathrm{7P_{3/2} \rightarrow 6D_{5/2}}$ transitions at  $\mathrm{15.57\; \mu m}$ and $\mathrm{14.59\;\mu m}$ respectively. The $\mathrm{8S}$, $\mathrm{8P}$, $\mathrm{4F}$ levels (grey color) are relevant for our numerical calculations. (b) Schematic of the experimental set-up. 
\label{Fig1}}
\end{figure}

A 459 nm or a 455 nm extended cavity laser diode pumps the cesium atoms to  the $\mathrm{7P_{1/2}}$ or $\mathrm{7P_{3/2}}$ level respectively. An auxiliary  saturated absorption set-up is also implemented in order to calibrate the frequency of the lasers. Before entering the cell, the beam passes through a pinhole after which it is collimated with a diameter of 3-4 mm. This creates a smooth beam profile without any intensity spikes. The maximum pump power entering the cell is $\mathrm{\approx1.5\; mW}$. The pump beam is also amplitude modulated (AM) using a chopper at frequencies of about 1 kHz.

The probe beam also comes from an extended cavity laser diode that can be tuned over several tens of nanometers around $\mathrm{1.53\; \mu m}$. The laser can be scanned continuously over several GHz with a very good scan frequency reprehensibility. The probe beam is also spatially filtered through a pinhole and collimated to a diameter of about 3 mm (slightly less than the pump beam diameter). 

The two lasers are superposed using a dichroic mirror before entering the cell. For the range of cesium densities explored here, the blue pump is absorbed in the body of the cell. For the highest cesium densities ($\mathrm{n_{Cs}\approx 8.10^{14}} \; \mathrm{cm^{-3}}$) the pump beam is almost fully absorbed within $\mathrm{\approx 100 \; \mu m}$ inside the cell. This means that near-field effects can be ignored as they are only important in the nanometric regime \cite{failache_prl_2002, Joao} for distances less than $\mathrm{1\;\mu m}$. Only for very low cesium densities (for $\mathrm{n_{Cs}}$ below $\mathrm{\approx 5.10^{12}\; cm^{-3}}$) the blue pump is not fully absorbed inside the 8cm long cell. For this purpose, at the other end of the cell the beams are separated using another dichroic mirror and their power is monitored by a silicon (blue light) or a germanium (infrared light) photodiode. Bandpass filters are also used to ensure that only light of the appropriate wavelength reaches our detectors. In all cases, the absorption of the infrared probe remains small, on the order of 10-100 ppm.

The principle of the experiments is the following: a strong blue laser at 459 nm  (455 nm) pumps atoms to the $\mathrm{7P_{1/2}}$ level ($\mathrm{7P_{3/2}}$). The pump laser selects atoms with a narrow class of velocities along the beam propagation axis. The velocity selection is much narrower than the Doppler width and in most cases is considered negligeable. BBR and inter-atomic collisions redistribute the atomic population to neighboring energy levels (Fig.2). The probe transmission spectrum, demodulated at the AM frequency of the pump, is then measured as a function of cesium density ($\mathrm{n_{Cs}}$) and cell temperature ($\mathrm{T_{cell}}$). The demodulated transmission is the difference between the pump-on ($P_{\mathrm{on}}$) and pump-off ($P_{\mathrm{off}}$) transmitted probe powers, divided by $P_{\mathrm{off}}$. Collisional process involve at least two atoms having nonlinear dependence on atomic density. As a general, but not absolute rule, collisions tend to redistribute the excitation in a broad distribution of atomic velocities \cite{gorlicki_prl_1982, van_kampen_pra_1997, huennekens_pra_1995}. In contrast, BBR radiation does not affect the atomic velocity (here the recoil kick can be ignored) and is a linear process with respect to the total number of atoms on the $\mathrm{7P_{1/2}}$ excited level.

\subsection*{BBR redistribution}


\begin{figure}[!htb]
\begin{center}
\includegraphics[width=80mm]{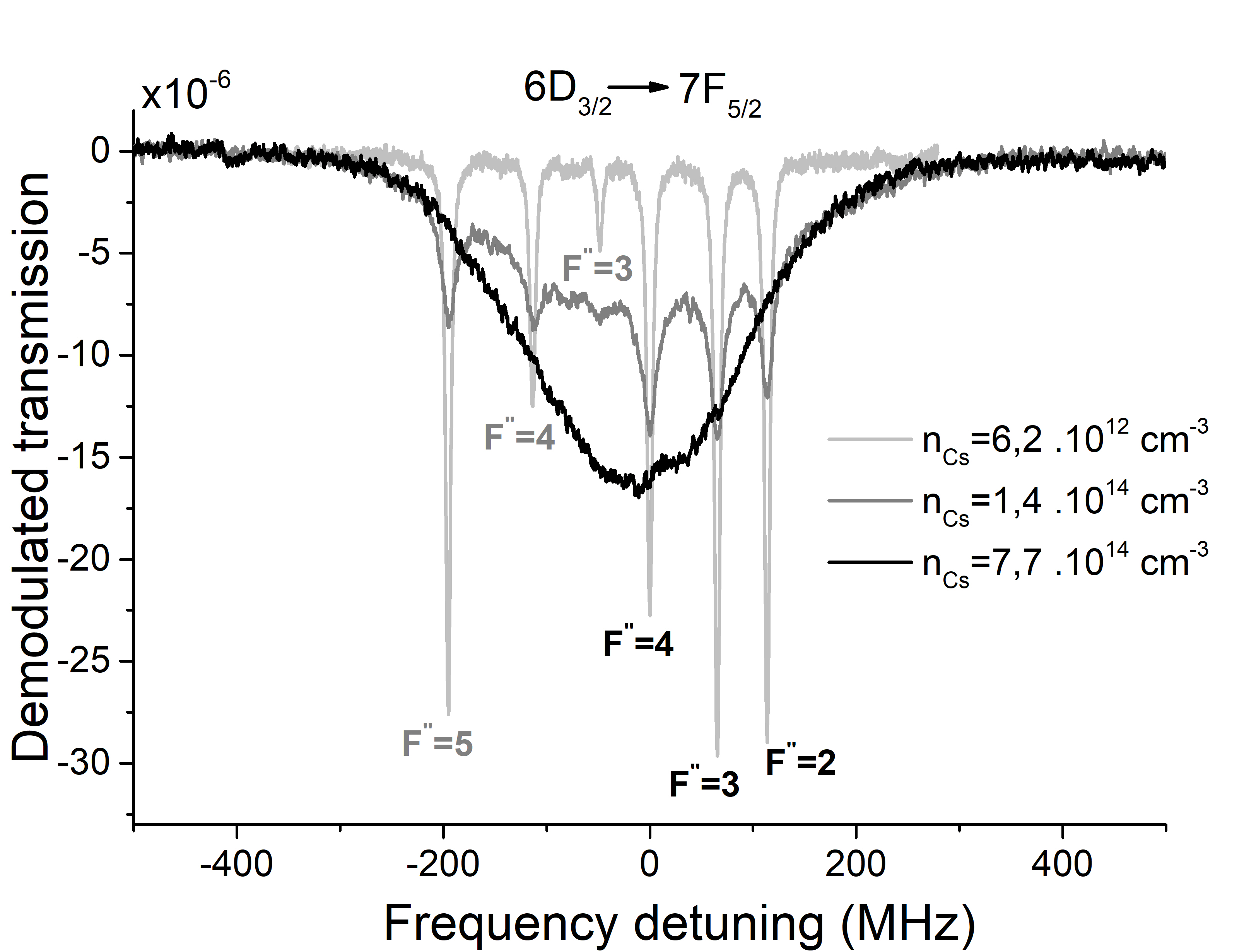}\\%
\end{center}
\caption{ (a) Transmission of the infrared probe laser (after demodulation) tuned at the  $\mathrm{6D_{3/2} \rightarrow 7F_{5/2}}$ transition as a function of the laser frequency for various cesium vapor densities and for $\mathrm{T_{cell} = 490 \; K}$. At low cesium densities sub-Doppler peaks are observed attributed to a $\mathrm{7P_{1/2} \rightarrow 6D_{3/2}}$ velocity preserving BBR absorption (see main text). The components $\mathrm{F^{''}=2,3,4}$ (black letters) correspond to atoms of $\mathrm{u_{\alpha} = 0\; m/s}$ whereas components $\mathrm{F^{''} = 3,4,5}$ (gray letters) correspond to atoms of ($\mathrm{u_{\beta} = -173 \; m/s}$). The scans are centered on the $\mathrm{F^{''}=4}$ ($\mathrm{u_{\alpha}=0\;m/s}$) hyperfine component.}
\label{Fig2}
\end{figure}

For the first set of experiments described here, the infra-red probe laser is scanned around the $\mathrm{6D\rightarrow 7F}$ transitions (see Fig.1a). Fig.2 shows the transmission spectrum of the probe laser for different atomic densities at a cell temperature of $\mathrm{T_{cell} = 490 \; K}$. Here, the pump laser is tuned on the $\mathrm{6S_{1/2} (F=4) \rightarrow 7P_{1/2} (F^{'}=3)}$  transition frequency, with a power of $\mathrm{\approx 0.7\; mW}$. 
The wavenumber of pump and probe laser is denoted by                   $\mathrm{k_{pump}}$, $\mathrm{k_{probe}}$ respectively. At low densities, the influence of collisions can be ignored. This suggests that ground state atoms with a velocity $\mathrm{u_{\alpha}}$ (here $\mathrm{u_{\alpha}=0}$) are pumped to the $\mathrm{F^{'}=3}$ hyperfine level of $\mathrm{7P_{1/2}}$, whereas atoms of velocity  $\mathrm{u_{\beta}}$, with $\mathrm{k_{pump} u_{\beta}=k_{pump} u_{\alpha}-2 \pi \Delta}$ (here $\mathrm{u_{\beta}=-173 \; m/s}$), are pumped to the $\mathrm{F^{'}=4}$ hyperfine level of $\mathrm{7P_{1/2}}$. Here $\mathrm{\Delta=377\; MHz}$ is the frequency spacing of the  hyperfine manifold, whereas $\mathrm{u_{\alpha}}$ and $\mathrm{u_{\beta}}$ are the velocity components along the beam propagation axis. BBR pumps the former velocity class to the $\mathrm{6D_{3/2} (F^{''}=2,3,4)}$ levels and the latter to  $\mathrm{6D_{3/2} (F^{''}=3,4,5)}$, which are probed by the infrared laser, scanned around the $\mathrm{6D_{3/2}\rightarrow 7F_{5/2}}$ transition, leading to the six peaks in Fig.2. When the pump laser is detuned by $\mathrm{\delta_{pump}}$ with respect to the $\mathrm{6S_{1/2}(F=4) \rightarrow 7P_{1/2}(F^{'}=3)}$  frequency then all six peaks shifts by $\mathrm{\delta_{pump}\frac{k_{probe}}{k_{pump}}}$ (for our experiments $\mathrm{\frac{k_{probe}}{k_{pump}} \approx 0.3}$) consistent with velocity selective pump probe spectroscopy. The observed linewidth of the peaks is limited to about 5MHz. This value does not depend on density and is similar for all the observed peaks, suggesting that it is probably related to the frequency instabilities of the lasers at timescales of a few seconds (time required to scan the frequency around a peak). The hyperfine manifold of the $\mathrm{7F_{5/2}}$ level is not resolved here, suggesting that hyperfine structure frequency spacing is well below 5MHz. This sets an upper limit for the magnetic dipole constant, $\mathrm{|A_{7F_{5/2}}|<0.5\; MHz}$, a value that seems consistent with previous dedicated measurements for $\mathrm{5F}$ and $\mathrm{6F}$ levels \cite{arimondo_revmodphys_1977, svanberg1975}.

For cesium densities lower than $\mathrm{\approx 10^{13} \; cm^{-3}}$ the effects of collisions are negligible and the transmission spectrum is independent of cesium density (see light gray curve in Fig.2). This is clear proof that in this regime collisional mechanisms are not involved in the observed $\mathrm{7P_{1/2} \rightarrow 6D_{3/2}}$ transfer. As the atomic density increases, one notices a broadening of the peaks as well as the appearance of a Doppler broadened background that eventually dominates the spectrum (Fig.2). Both these effects can be attributed to collisions.

\begin{figure}[!htb]
\begin{center}
\includegraphics[width=80mm]{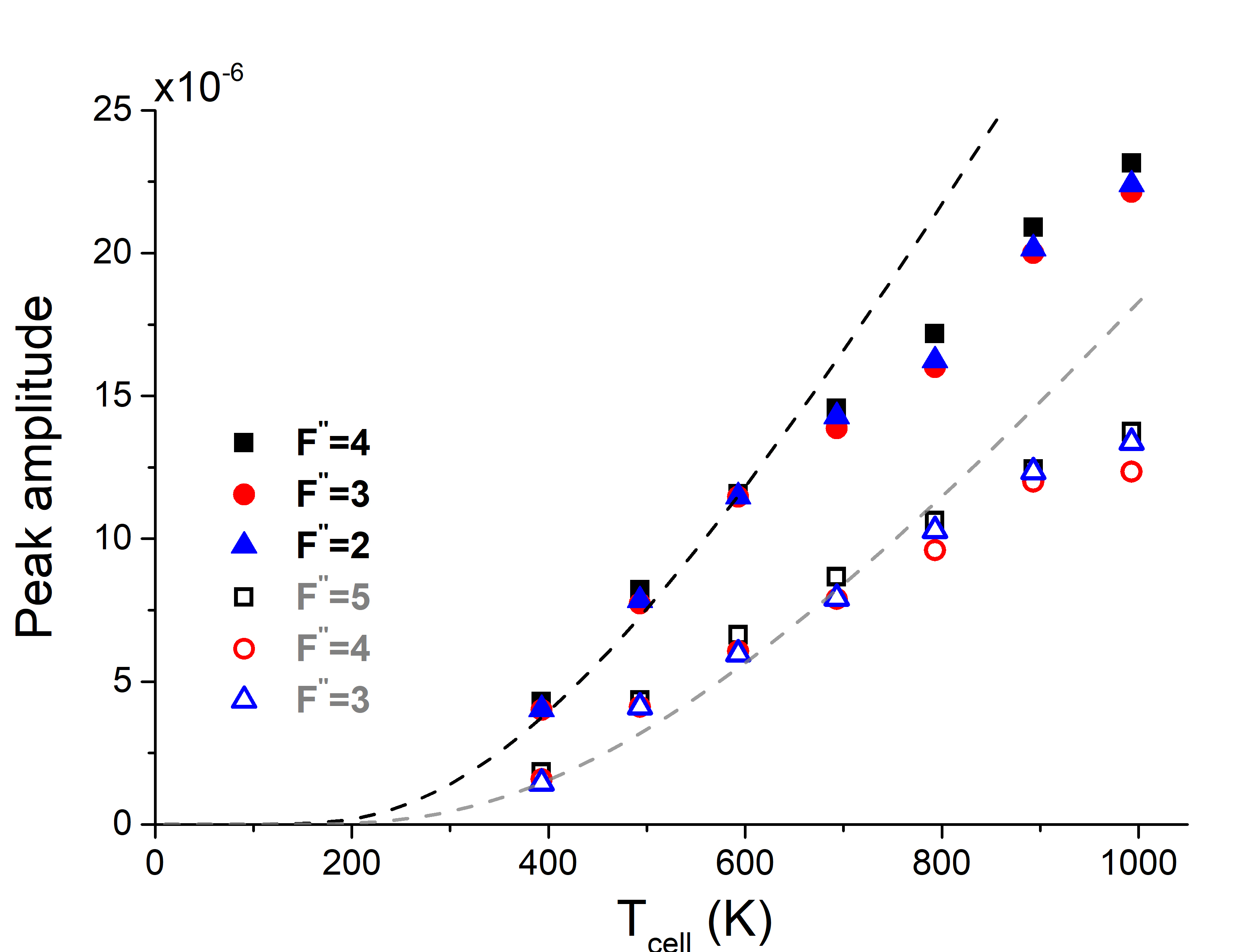}\\%
\end{center}
\caption{Amplitude of all the sub-Doppler peaks observed at the $\mathrm{6D_{3/2}\rightarrow7F_{5/2}}$ transition, divided (normalized) by the relative weights of the $\mathrm{7P_{1/2} (F^{'}) \rightarrow 6D_{3/2}(F^{''})} $ hyperfine transitions. Filled points correspond to transitions starting from $\mathrm{F^{''}=2,3,4}$ levels (black letters) with $\mathrm{u_{\alpha} = 0\; m/s}$ whereas open points to $\mathrm{F^{''}=3,4,5}$ levels (gray letters) with $\mathrm{u_{\beta} = -173\; m/s}$. The error bar associated with the $\mathrm{T_{cell}}$ measurement is $\approx$ 30 K. The dashed curves represent the expected evolution according to the Bose-Einstein factor. The reservoir temperature is fixed at $\mathrm{T_r}$=370 K. The cesium density varies with $\mathrm{T_{cell}}$ approximately from $\mathrm{1.4 \cdot 10^{13} cm^{-3}}$ to $\mathrm{5.8 \cdot 10^{12} cm^{-3}}$. }
\label{Fig3}
\end{figure}

In Fig.3 we plot the amplitude of the sub-Doppler peaks as a function of cell temperature, while keeping the vapor density at moderate levels and therefore ensuring that collisional transfer to the $\mathrm{6D_{3/2}}$ level is negligible. We show the amplitudes of transitions starting from $\mathrm{6D_{3/2} (F^{''}=2,3,4)}$ levels (for atoms with velocity $\mathrm{u_{\alpha}}$=0 m/s) as well as transitions from $\mathrm{6D_{3/2} (F^{''}=3,4,5)}$ levels (for atoms with velocity $\mathrm{u_{\beta}}$=-173 m/s). The amplitudes are divided (normalized) by the relative weights of the $\mathrm{7P_{1/2} (F^{'}) \rightarrow 6D_{3/2}(F^{''}) }$ hyperfine transitions. These are $\mathrm{(15/56,21/56,20/56)} $ and  $\mathrm{(7/72,21/72,44/72) }$ for the   $\mathrm{F^{'}=3 \rightarrow F^{''}=2,3,4} $  and $\mathrm{F^{'}=4 \rightarrow F^{''}=3,4,5 }$ transition respectively. Transitions starting from $\mathrm{F^{''}=3,4,5}$ levels are smaller by a factor corresponding to the population ratio of the two velocities multiplied by the strength ratio between the $\mathrm{6S_{1/2} (F=4) \rightarrow 7P_{1/2}(F^{'}=3,4)} $ transitions. For temperatures below 700 K (in this range the cell body temperature is homogeneous), the amplitude evolution of all peaks follows Bose-Einstein statistics given by $\mathrm{n(\lambda,T_{cell})=[ e^{\frac{hc  }{(k_B T_{cell})\lambda}}-1]^{-1}} $, where $\mathrm{h}$ is Planck's constant, $\mathrm{c}$ the speed of light and $\mathrm{\lambda}$ the wavelength. This demonstrates that the population transfer is due to BBR. As aforementioned, when $\mathrm{T_{cell}}$ is higher than 700 K, the temperature of the second window, $\mathrm{T_{inter}}$, stays fixed at 700 K. In this case the temperature of the cell body is inhomogeneous. For this reason, the experimental points in Fig. 3 fall below the theoretical expectations that do not account for a temperature gradient of the cell temperature.

\begin{figure}[!htb]
\begin{center}
\includegraphics[width=80mm]{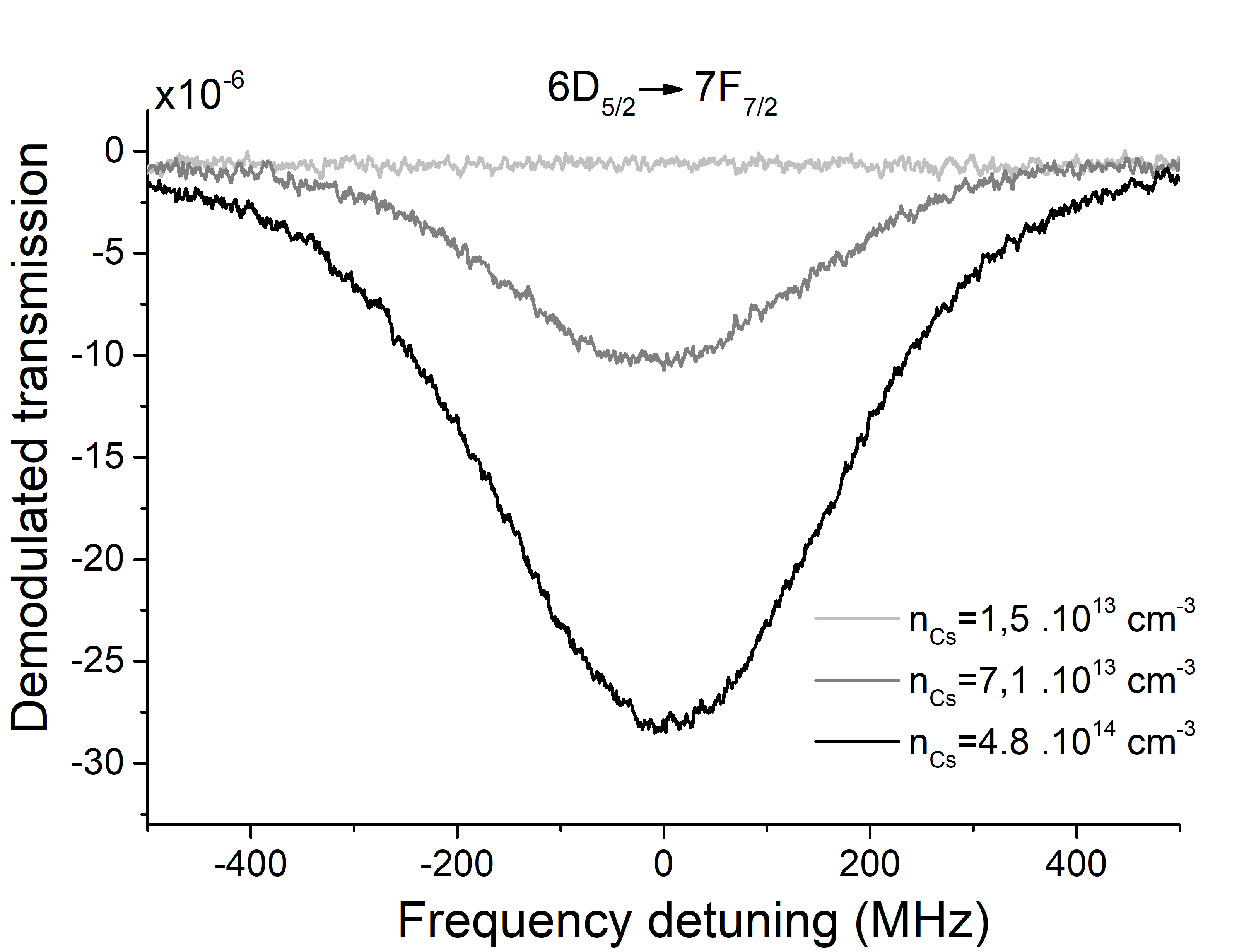}\\%
\end{center}
\caption{Probe transmission (after demodulation) at the  $\mathrm{6D_{5/2} \rightarrow 7F_{7/2}}$ transition for various cesium densities at $\mathrm{T_{cell}}$=800 K. The scans are centered on the peak of the spectrum. The FWHM spread of the distribution is  $\mathrm{\approx }$350MHz for $\mathrm{n_{Cs}=1.4 \cdot 10^{14} cm^{-3}}$ and $\mathrm{\approx }$390MHz for $\mathrm{n_{Cs}=8.3 \cdot 10^{14} cm^{-3}}$. The hyperfine structure spread of the $\mathrm{6D_{5/2}}$ is $\mathrm{\approx }$90MHz \cite{georgiades_optlett_1994, arimondo_revmodphys_1977}, whereas the FWHM Doppler width of a single transition at this temperature is 345MHz.
\label{Fig4}}
\end{figure}

Furthermore, we have conducted experiments on the $\mathrm{6D_{5/2} \rightarrow 7F_{7/2}}$ transitions, shown in Fig. 4. The $\mathrm{6D_{5/2}}$ level cannot be directly populated by BBR as the $\mathrm{7P_{1/2}\rightarrow6D_{5/2}}$ transition is dipole forbidden. At high vapor densities collisional mechanisms distribute the excitation from the $\mathrm{7P_{1/2}}$ to the $\mathrm{6D_{5/2}}$ level. A broad spectrum is observed suggesting that the excitation is distributed to a wide range of velocities. It should be noted that pump power and frequency were the same for the experiments of Fig.2 and Fig.4. The two figures are plotted on the same vertical scale for a direct comparison between transfer mechanisms. 

Experiments were also performed using a 455 nm laser  pumping the atoms to the $\mathrm{7P_{3/2}}$ level and a probe laser scanning around the $\mathrm{6D\rightarrow7F}$ transitions. In this case, the signals are also sub-Doppler but significantly more complicated due to the hyperfine manifold of the $\mathrm{7P_{3/2}}$ levels.

\subsubsection*{Discussion}

Our studies of BBR excitation redistribution focus primarily on the $7P\rightarrow 6D$ channel. In principle BBR can transfer the initial excitation to many adjacent energy levels with transfer rate proportional to the Bose-Einstein factor at a given transition wavelength. To estimate the population of energy levels surrounding the $7P$ pumping level we have solved the system of rate equations (see \cite{pimbert_1972} for more details) in the low density and low power limit when collisions and saturation can be safely ignored. Our calculations show that the excitation is mainly distributed to the $\mathrm{6D}$, $\mathrm{8S}$ and $\mathrm{7D}$ levels (see Fig.1a) by BBR transfer and to lower lying states , such as the $\mathrm{7S}$ and $\mathrm{5D}$, by both spontaneous emission and BBR transfer (both preserving the atomic velocities \cite{zhou_cs_2017}). Interestingly, our calculations show that for high temperatures the $\mathrm{8P}$ and $\mathrm{4F}$ levels can also be significantly populated due to two-step BBR excitations. For example, at $\mathrm{T_{cell}}$=1000 K, $\mathrm{8P}$ and $\mathrm{4F}$ populations are about $\mathrm{6.5\%}$ and $\mathrm{1.3\%}$ of the $\mathrm{6D_{3/2}}$ population (probed in our experiments) whereas the $\mathrm{8S}$  population is about $\mathrm{80\%}$ of the $\mathrm{6D_{3/2}}$ population. This suggests that using a similar set-up, one could use the $\mathrm{4F\rightarrow nG}$ transitions to perform high-resolution, sub-Doppler spectroscopy of $\mathrm{nG_{7/2}}$ and $\mathrm{nG_{9/2}}$ levels that are normally accessible via quadrupole-quadrupole spectroscopy from the ground state \cite{weber_pra_1987}. Similarly, spectroscopy of higher nP levels can be performed on the $\mathrm{8S\rightarrow nP}$ transitions that also lie in infrared wavelengths. However, for these experiments the available signal will be significantly reduced.  

\subsection*{Collisional redistribution}

While the collisional redistribution within the $6P$ level has been experimentally studied in the past \cite{huennekens_pra_1995, failacheprl1999, failacheEPJD2003}, similar studies for high-lying states have been scarce. In order to study collisional redistribution processes within the $\mathrm{7P}$ manifold we turn our infrared laser on resonance with the $\mathrm{7P_{1/2} \rightarrow 10S_{1/2}}$ transition at 1530 nm. In these experiments the beam waist of the lasers (pump and probe) was reduced ($\mathrm{\approx 500\mu m}$) in order to increase the available pump intensity. This allows us to achieve intensities significantly higher than the saturation intensities of the $\mathrm{7P}$ levels ($\mathrm{\approx}$60 $mW$/$cm^2$ and $\approx$15 $mW$/$cm^2$ for the $\mathrm{7P_{1/2}}$ and the $\mathrm{7P_{3/2}}$ respectively \cite{antypas_pra_2013}). In the first experiment, we probe the $\mathrm{7P_{1/2} \rightarrow 10S_{1/2}}$ transitions while pumping directly the atoms to the $\mathrm{7P_{1/2}}$ level. This allows us to study the redistribution of the excitation within the hyperfine manifold of the $\mathrm{7P_{1/2}}$ level. 

In Fig.5a we plot the infrared probe transmission through the cell for three different cesium densities and for a cell temperature of $\mathrm{T_{cell}=570K}$. The 459 nm laser pumps atoms of $\mathrm{u_{\alpha}=0 m/s}$ and $\mathrm{u_{\beta}=-173 m/s}$ to the $\mathrm{F^{'}=3}$ and $\mathrm{F^{'}=4}$ hyperfine levels of $\mathrm{7P_{1/2}}$ respectively. These atoms are probed by the infrared laser, leading to the four main peaks of the infrared transmission spectrum (Fig.5). $\mathrm{F^{'}=3\rightarrow F^{''}=3,4}$ (for $\mathrm{u_{\alpha}}$ atoms) at  0 MHz and -252 MHz and $\mathrm{F^{'}=4\rightarrow F^{''}=3,4}$ (for $\mathrm{u_{\beta}}$ atoms) at -742 MHz and -490 MHz. At very low cesium densities (light gray curve) these are the only observed peaks. Strikingly, additional sub-Doppler peaks appear as cesium density increases. These peaks seem to correspond to a collisional transfer of $\mathrm{u_{\beta}}$ and $\mathrm{u_{\alpha}}$ atoms to the $\mathrm{F^{'}=3}$ and $\mathrm{F^{'}=4}$ levels respectively. This should lead to four additional peaks: $\mathrm{F^{'}=3\rightarrow F^{''}=3,4}$, for $\mathrm{u_{\beta}}$ atoms at $\mathrm{-113}$ MHz and $\mathrm{-365}$ MHz (shifted by $\mathrm{-\Delta\frac{k_{pump}}{k_{probe}}\approx-113}$ MHz with respect to those observed for $\mathrm{u_{\alpha}}$ atoms) and $\mathrm{F^{'}=4\rightarrow F^{''}=3,4}$, for $\mathrm{u_{\alpha}}$ atoms at -629 MHz and -377 MHz (shifted by $\mathrm{\approx+113}$ MHz with respect to those observed for $\mathrm{u_{\beta}}$ atoms). The above hypothesis is consistent with the observations as the -377 MHz and -365 MHz peaks partly overlap. Although a broad collisional background also appears, the sub-Doppler peaks remain visible for very high densities. It is worth noting that the width of the additional peaks is similar to the that of the main peaks (corresponding to the atoms directly pumped by the 459 nm laser). 

Fig5a shows that the additional sub-Doppler peaks depend on cesium density. This demonstrates that the peaks are due to a collisional transfer and not due to two step BBR processes that could in principle redistribute the excitation within the hyperfine levels while preserving the atomic velocities. Furthermore, we have used the previously mentioned rate equation model to calculate the population redistribution between the hyperfine components of $\mathrm{7P_{1/2}}$ due to two-step BBR processes. According to our findings, for a cell temperature of $\mathrm{T_{cell}}$=570K two-step BBR processes are negligible and cannot account for the findings of Fig5. 

\begin{figure}[htb]
\begin{center}
\includegraphics[width=80mm]{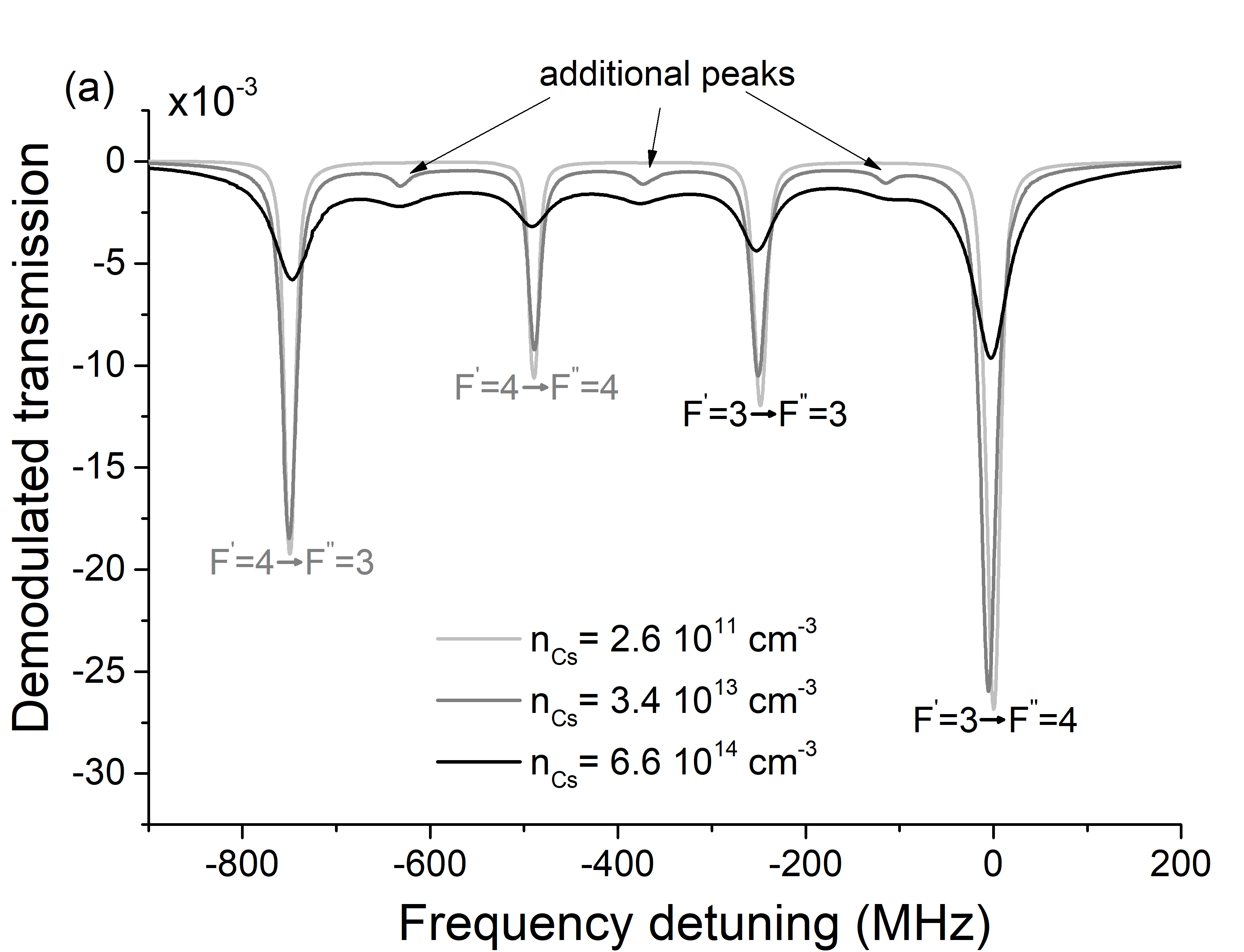}\\%
\includegraphics[width=80mm]{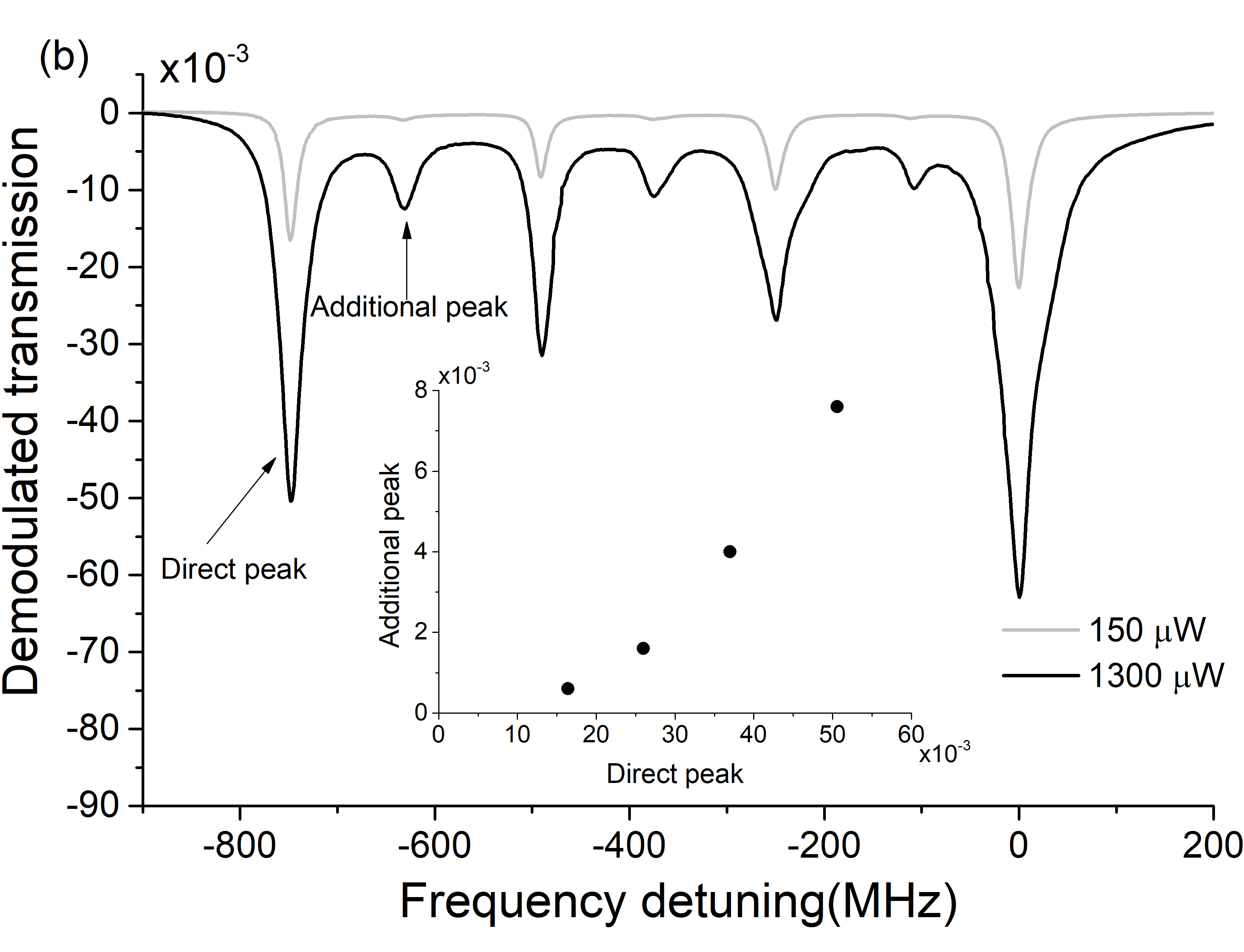}%
\end{center}
\caption{ Probe transmission (after demodulation) of the infrared probe laser tuned at the $\mathrm{7P_{1/2}(F^{'}) \rightarrow 10S_{1/2} (F^{''})}$ transition. (a) Transmission spectrum for different cesium densities. The 459 nm pump laser power is 200 $\mathrm{\mu}$W and the cell temperature is $\mathrm{T_{cell}}$=570 K. The four main (directly pumped) peaks are denoted on the graph $\mathrm{F^{'}=3\rightarrow F^{''}=3,4}$ for atoms with velocity $\mathrm{u_{\alpha}}$ (black letters) and $\mathrm{F^{'}=4\rightarrow F^{''}=3,4}$ for atoms with velocity $\mathrm{u_{\beta}}$ (gray letters). Additional sub-Doppler peaks appear when cesium density increases (b) Transmission spectrum for different pump powers with $\mathrm{n_{Cs}}$=8.3 $\mathrm{10^{12}}$ $cm^{-3}$ and $\mathrm{T_{cell}}$=570 K. The inset shows the amplitude of the additional peak (indicated on the graph) as a function of the amplitude of the direct peak (indicated on the graph) for different pump powers (150 $\mathrm{\mu}$W, 300 $\mathrm{\mu}$W, 700 $\mathrm{\mu}$W and 1300 $\mathrm{\mu}$W). }
\label{Fig5}
\end{figure}

In Fig5b we show the the $\mathrm{7P_{1/2} \rightarrow 10S_{1/2}}$ transmission spectrum for two different pump powers (150 $\mathrm{\mu}$W and 1300 $\mathrm{\mu}$W). We can observe that the amplitude evolution of the direct peaks increases sub-linearly with pump power verifying that the pump laser saturates the 459 nm transition. More importantly, we observe that the additional sub-Doppler peaks increase much faster with pump power compared to the main peaks while the width of direct and additional peaks remains comparable. The amplitude of one additional peak is plotted as a function of the amplitude of one direct peak for four different pump powers (150 $\mathrm{\mu}$W, 300 $\mathrm{\mu}$W, 700 $\mathrm{\mu}$W and 1300 $\mathrm{\mu}$W) in the inset of Fig.5b. The points follow a super-linear trend. The above observations are a strong indication that the collisional mechanism, giving rise to the additional sub-Doppler peaks, involves collisions between two excited atoms. Such collisional mechanisms should depend quadraticaly on the excited state population (see Discussion section below). Here, the amplitude of the direct peaks depends on the excited state population directly pumped to the $\mathrm{7P}$ level, while the amplitude of the additional peaks depends on the collisionally transferred population. This justifies the super-linear dependence of the additional peak amplitude as a function of the direct peak amplitude, shown in inset of Fig5b.   

Surprisingly, the additional sub-Doppler peaks correspond to a collisional but velocity preserving transfer between the hyperfine manifold of the $\mathrm{7P_{1/2}}$ level. Additionally, our experimental findings strongly suggest that this collisional mechanism involves two velocity selected excited atoms. This observation is in direct opposition to the collisional redistribution measured within the $\mathrm{6P_{1/2}}$ hyperfine manifold that shows little or no evidence of velocity preservation \cite{huennekens_pra_1995}. 

In the second experiment we pump the atoms to the $\mathrm{7P_{3/2}}$ level while still probing the population of the $\mathrm{7P_{1/2}}$ level. In Fig. 6 we show the probe transmission spectrum at the $\mathrm{7P_{1/2} \rightarrow 10S_{1/2}}$ transition for two different pump laser powers. Here $\mathrm{T_{cell}}$=490 K and $\mathrm{n_{Cs}}$=8.3 $\mathrm{10^{13}}$ $\mathrm{cm^{-3}}$. The $\mathrm{7P_{1/2}}$ level is populated by collisions that redistribute atoms within the $\mathrm{7P}$ fine structure. As in the previous case the transmission spectrum clearly displays evidence of velocity selection. The peaks observed in the spectrum are here roughly 70 MHz broad but significantly narrower than the Doppler FWHM linewidth of $\approx$ 270 MHz. It should be noted that pump laser selects three velocities, one for each hyperfine transition $\mathrm{F=4 \rightarrow F^{'}=3,4,5}$. Here, the hyperfine spacing $\mathrm{F^{'}=3,4,5}$ is roughly 150 MHz (see Fig.1a). In the infrared spectrum the Doppler shift between the velocity components is multiplied by $\mathrm{\frac{k_{probe}}{k_{pump}}}$ giving $\mathrm{\approx}$ 50MHz. This could partly explain the 70 MHz observed linewidth of the peaks. 

\begin{figure}[!htb]
\begin{center}
\includegraphics[width=80mm]{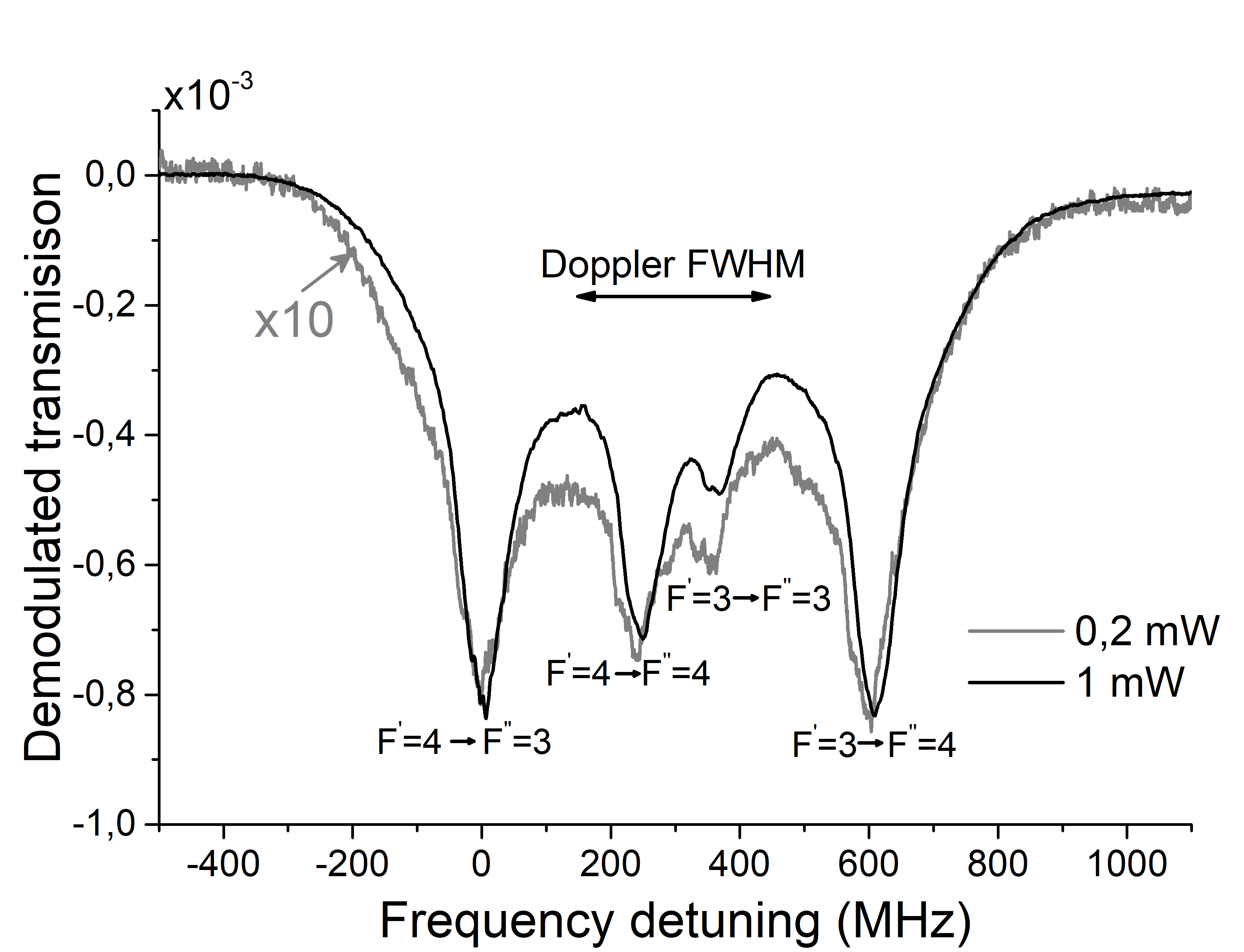}\\%
\end{center}
\caption{ Probe transmission (after demodulation) on the $\mathrm{7P_{1/2} \rightarrow 10S_{1/2}}$ transition at 1530 nm when the pump is tuned on the $\mathrm{6S_{1/2}(F=4)\rightarrow 7P_{3/2}(F^{'}=5)}$ transition. Here $\mathrm{T_{cell}}$=490 K and $\mathrm{n_{Cs}}$=6.8 $\mathrm{10^{13}}$ $\mathrm{cm^{-3}}$. The $\mathrm{7P_{1/2}}$ level is mainly populated by collisions. The scans are arbitrarily centered on the peak of smallest frequency. The Doppler FWHM width is shown for comparison. The transmission spectrum is measured for two different pump powers 1 mW (black line) and 0.2 mW (gray line). The gray curve is multiplied by a factor of 10. The lineshapes of the two curves are not the same indicating an interplay between linear and super linear processes. }
\label{Fig5}
\end{figure}

In Fig.6 we can see that the transmission change is superlinear with pump power since an increase of the pump power by a factor of 5 leads to an increase of signal approximately by a factor of 10. This suggests that fine structure redistribution is also due to collisions between two excited cesium atoms transferring population to the $\mathrm{7P_{1/2}}$ level. This mechanism would lead to a quadratic dependence of the transmission signal with $\mathrm{7P_{3/2}}$ population (that is defined by pump power). Nevertheless, in the conditions of our experiment it is difficult to predict the exact dependence of the transmission spectrum as a function of pump power because: (a) collisional transfer to the Cs($\mathrm{7P_{1/2}}$) can also be achieved via collisional processes between Cs($\mathrm{7P}$) and Cs($\mathrm{6S}$) atoms \cite{pace_canjp_1974, Chevrollier_OL_91}, which should lead to a broad background varying linearly with $\mathrm{7P_{3/2}}$ population (b) saturation of the pump transition leading to a sub-linear dependence of the transmission spectrum with pump power cannot be ignored as the saturation intensity of the pump transition is $\approx$ 15 $\mathrm{mW}$/$\mathrm{cm^2}$.  

\subsubsection*{Discussion}

We now discuss the underlying collisional mechanisms that could explain our experimental observations. Exchange collisions between $\mathrm{7P}$ and $\mathrm{6S}$ are expected to redistribute the atomic population and atomic velocities within the fine and hyperfine manifold of the $\mathrm{7P}$ state. The kernel of exchange collisions \cite{gorlicki_prl_1982, berman_pra_1991} is broad with a small memory of the initial velocity leading to fast thermalization. The number of such collisions per unit time and unit volume is proportional to $\mathrm{\sim k_{G} \cdot n_{G} \cdot n_{E}}$, where $\mathrm{k_{G}}$ is the collision rate coefficient of the process, $\mathrm{n_G}$ is the ground state population (approximately  equal the total cesium density) and $\mathrm{n_E}$ is the excited state ($\mathrm{7P}$) population. This process is therefore sub-linear with pump power (linear in the weak pumping regime). These collisional mechanisms could explain the Doppler broadened background observed in Fig.5 and Fig.6 but are unable to explain the behavior of the observed sub-Doppler contributions.
     
Instead the sub-Doppler contributions seem to originate from collisions between excited state atoms with velocity components $\mathrm{u_{\alpha}}$ and $\mathrm{u_{\beta}}$, selected by the blue pump laser. Velocity selection is also preserved after spontaneous emission to the downward $\mathrm{6D}$ and $\mathrm{7S}$ levels (see Fig.1a) \cite{zhou_cs_2017} but is probably lost after further decay to the $\mathrm{6P}$ level \cite{huennekens_pra_1995}. For collisional processes between excited state atoms the number of collisions per unit time per unit volume is proportional to  $\mathrm{\sim k_E \cdot n^1_{E} \cdot n^2_{E}}$, with $\mathrm{k_E}$ the collision rate coefficient and $\mathrm{n^1_{E}}$, $n^2_{E}$ the population of the two excited states involved in the collision. These processes can be super-linear with pump power and are consistent with the observations of Fig. 5 and Fig. 6.

One possible mechanism that can lead to hyperfine structure redistribution is resonant exchange collisions between two velocity selected atoms of $\mathrm{7P}$ and $\mathrm{5D}$ or $\mathrm{7S}$ levels, both significantly populated due to spontaneous emission ($\mathrm{n_{5D}\approx1.5n_{7P}}$ while $\mathrm{n_{7S}\approx0.2n_{7P}}$).  This would be according to the process:\\
\begin{multline}
   \mathrm{Cs(7P_{1/2},F,u_{\alpha})+Cs(5D_{3/2},F^{\prime},u_{\beta})}\rightarrow 
    \\
    \mathrm{Cs(7P_{1/2},F,u_{\beta})+Cs(5D_{3/2},F^{\prime},u_{\alpha})}
\end{multline}
\\

The above process does not require deflection of velocities and therefore the selected velocity components (along the pump propagation axis) $\mathrm{u_{\alpha}}$ and $\mathrm{u_{\beta}}$ can be preserved \cite{huennekens_pra_1995}. The collision rate per unit volume of such resonant exchange collisions should depend quadraticaly on the excited ($\mathrm{7P}$ population), assuming that spontaneous emission stays the dominant population mechanism of the $\mathrm{5D}$ or $\mathrm{7S}$ levels.   

Finally, we note that fine structure redistribution could be possibly due to a different process as it also involves a significant change of the internal energy of the products. Our experiments show that fine structure redistribution is less efficient (the signal amplitude of Fig.6 is smaller than that of Fig.5 for similar pumping strength) causing additionally significant broadening of the initial velocity selection.

\section{Conclusions }
Our experiments study the redistribution of an atomic population, initially pumped to the second cesium resonance, to many adjacent energy levels. At high cesium densities, this redistribution is mainly due to collisions. We show that collisional redistribution within the $\mathrm{7P}$ levels can happen via fine structure changing and hyperfine structure changing collisions between excited state atoms that preserve the atomic velocities. We observe also that exchange collisions and radiation trapping are not effective in redistributing (thermalizing) the excitation in the $\mathrm{7P}$ state. This is in sharp contrast with observations performed for the cesium $\mathrm{6P}$ state, where the initial velocity selection of the pump is almost lost even at very low densities \cite{huennekens_pra_1995}. State changing collisions preserving atomic velocities could become even more prominent for higher lying excited atomic states and Rydberg atoms \cite{Gallagher_1988} where the importance of resonant exchange collisions with ground state atoms is expected to diminish. In this respect, the collisional mechanisms observed here could play a role in measurements of collective effects between Rydberg atoms in thermal vapor cells \cite{kubler_natphoton_2010, Urvoy_PRL_2015} as well as in measurements of Rydberg population \cite{barredo_prl_2013} or buffer gas density \cite{Schmidt_2020} using Rydberg ionization via collisions or to a lesser extent by BBR \cite{Beterov_NJP_2009}. Additionally, our observations can be of importance for two-step Rydberg spectroscopy via the second instead of the first atomic resonance.    

At low cesium densities we observe velocity preserving redistribution of the initial laser excitation due to absorption of BBR photons. Our experiments are performed in the volume of the cell where broadband BBR (Planck spectrum) significantly populates a plethora of energy levels at high temperatures. This velocity preserving process can provide a simple way of performing high-resolution spectroscopy of highly excited energy levels of alkali atoms. We also discuss the possibility of exploiting two step BBR processes to perform sub-Doppler spectroscopy of cesium $\mathrm{nG_{7/2}}$ and $\mathrm{nG_{9/2}}$ states. Finally, we mention that BBR transfer can be also important in the study of low lying states such as $\mathrm{6P}$ cesium, in particular on the $\mathrm{6P\rightarrow5D}$ or $\mathrm{6P\rightarrow7S}$ channels.  

An interesting perspective whould be to study the near-field $\mathrm{7P_{1/2} \rightarrow 6D_{3/2}}$ transfer due to thermally excited sapphire polaritons \cite{failacheprl1999, Joao}. This can be achieved by probing the $\mathrm{6D_{3/2} \rightarrow 7F_{5/2}}$ transition near the sapphire surface using selective reflection spectroscopy \cite{failache_prl_2002}. The signature of this near field energy transfer mechanism needs to be discriminated from collisional or far-field BBR transfer, studied in  this paper, that affect atoms in the volume of the cell.

\section*{Acknowledgments}
We would like to thank Jose Roberto Rios Leite, Maki Tachikawa and Martial Ducloy for interesting and fruitful discussions. We are grateful to Paul Berman for his comments on this manuscript.  We would also like to thank ECOS-Sud (ECOS U14E01) and the LIA, 'Institut Franco-Uruguayen de Physique' for financial support. J.C.A.C. acknowledges financial support from CAPES, Brazil.

\bibliography{biblioBBRpaper}


\end{document}